\documentstyle[12pt,aasms4]{article}   

\lefthead{Aloy et al.}
\righthead{HIGH--RESOLUTION 3D SIMULATIONS OF RELATIVISTIC JETS}

\def\be{\begin{equation}}
\def\ee{\end{equation}}
\newcommand\cref[1]{(\ref{#1})}

 \def\etal{{\sl et al.}\,}
 \def\ie{{\sl i.e.,}\,}
 \def\eg{{\sl e.g.,}\,}

 \def\la{\hbox{\raise.5ex\hbox{$<$}
     \kern-1.1em\lower.5ex\hbox{$\sim$}}}
 \def\ga{\hbox{\raise.5ex\hbox{$>$}
     \kern-1.1em\lower.5ex\hbox{$\sim$}}}

\begin{document}

\title{HIGH--RESOLUTION 3D SIMULATIONS OF RELATIVISTIC JETS}

\author{M.A. Aloy \altaffilmark{1}, 
        J.M$^{\underline{\mbox{a}}}$ Ib\'a\~nez \altaffilmark{1},
        J.M$^{\underline{\mbox{a}}}$ Mart\'{\i} \altaffilmark{1},
        J.L. G\'omez \altaffilmark{2}  and
        E. M\"uller \altaffilmark{3}
}

\altaffiltext{1} {Departamento de Astronom\'{\i}a y Astrof\'{\i}sica,
                  Universidad de Valencia, 46100 Burjassot (Valencia), Spain}

\altaffiltext{2} {Instituto de Astrof\'{\i}sica de Andaluc\'{\i}a (CSIC)
                  Granada, Spain}

\altaffiltext{3} {Max--Planck--Institut f\"ur Astrophysik,
                  Karl--Schwarzschild--Str.\,1, 85748 Garching, Germany}

\begin{abstract}
  We have performed high-resolution 3D simulations of relativistic
jets with beam flow Lorentz factors up to 7, a spatial resolution of 8
cells per beam radius, and for up to 75 normalized time units to study
the morphology and dynamics of 3D relativistic jets. Our simulations
show that the coherent fast backflows found in axisymmetric models are
not present in 3D models. We further find that when the jet is exposed
to non-axisymmetric perturbations, (i) it does not display the strong
perturbations found for 3D classical hydrodynamic and MHD jets (at
least during the period of time covered by our simulations), and (ii)
it does propagate according to the 1D estimate.  Small 3D effects in
the relativistic beam give rise to a lumpy distribution of apparent
speeds like that observed in M87. The beam is surrounded by a boundary
layer of high specific internal energy. The properties of this
layer are briefly discussed.
\end{abstract}


\section{Introduction } 
\label{s:intro}

  Since several years the dynamical and morphological properties of
axisymmetric relativistic jets are investigated by means of
relativistic hydrodynamic simulations (van Putten 1993; Duncan \&
Hughes 1994; Mart\'{\i} \etal 1994, 1995, 1997; Komissarov \& Falle
1998; Rosen \etal 1999). In addition, relativistic MHD simulations
have been performed in 2D (Koide, Nishikawa \& Muttel 1996; Koide
1997) and 3D (Nishikawa \etal 1997, 1998). In their 3D simulations
Nishikawa \etal have studied mildly relativistic jets (Lorentz factor
4.56) propagating both along and obliquely to an ambient magnetic
field. In this Letter we report on high-resolution 3D simulations 
of relativistic jets with the largest beam flow Lorentz factor performed
up to now (7.09), the largest resolution (8 cells per beam radius),
and covering the longest time evolution (75 normalized time units; a
normalized time unit is defined as the time needed for the jet to
cross a unit length; see Massaglia, Bodo \& Ferrari 1996). 

  The calculations have been performed with the high--resolution 3D
relativistic hydrodynamics code GENESIS (Aloy \etal 1999), which is
an upgraded version of the code developed by Mart\'{\i}, M\"uller \&
Ib\'a\~nez (1994) and Mart\'{\i} \etal (1995). GENESIS integrates the
3D relativistic hydrodynamic equations in conservation form in
Cartesian coordinates including an additional conservation equation
for the density of beam material.  The computations were performed on a
Cartesian domain (X,Y,Z) of size $15R_b \times 15R_b \times 75R_b$ 
($120 \times 120 \times 600$ computational cells), where $R_b$ is the 
beam radius. The jet is injected at $z=0$ along the 
positive $z$-axis through a circular nozzle defined by $x^2+y^2 \le R_b^2$. 
Beam material is injected with a beam mass fraction $f=1$, and the 
computational domain is initially filled with an external medium ($f=0$). 

  We have considered a 3D model corresponding to model C2 of
Mart\'{\i} \etal (1997), which is characterized by a beam-to-external
proper rest-mass density ratio $\eta=0.01$, a beam Mach number
$M_b=6.0$, and a beam flow speed $v_b = 0.99c$ ($c$ is the speed of
light) or a beam Lorentz factor $W_b \approx 7.09$. An ideal gas
equation of state with an adiabatic exponent $\gamma =5/3$ is assumed
to describe both the jet matter and the ambient gas. The beam is
assumed to be in pressure equilibrium with the ambient medium.  

  The evolution of the jet was simulated up to $T \approx 150 R_b/c$,
when the head of the jet is about to leave the grid. The mean jet
propagation speed $v_h \approx 0.5c$, while the 1D estimate of the jet
propagation speed (see, \eg Mart\'{\i} \etal 1997) gives $0.42c$, \ie
our simulations are still within the 1D phase (see Mart\'{\i},
M\"uller \& Ib\'a\~nez 1998). 

  Non--axisymmetry was imposed by means of a helical velocity
perturbation at the nozzle given by
\begin{equation}
  v^x_b = \zeta v_b \cos \left(\frac{2 \pi t}{\tau}\right), \;
  v^y_b = \zeta v_b \sin \left(\frac{2 \pi t}{\tau}\right), \;
  v^z_b = v_b \sqrt{1-\zeta^2},
\end{equation}
where $\zeta$ is the ratio of the toroidal to total velocity and $\tau$
the perturbation period (\ie $\tau = T/n$, $n$ being the number of
cycles completed during the whole simulation). The wavelength of the 
perturbation, $\lambda$, is obtained from the expression 
$\displaystyle{\lambda = v^z_b \tau \approx \frac{v_b}{v_h} \frac{L}{n}}$, 
where $L$ is the axial dimension of the grid.  

\section{Morphology and dynamics of 3D relativistic jets}
\label{s:morpho}

  We have considered a model with a $1\%$ perturbation in helical velocity 
($\zeta=0.01$) and $n=50$. Figure\,\ref{f:n50p01a} shows various quantities of 
the jet in the plane $y=0$ at the end of the simulation. Two values of the beam 
mass fraction are marked by white contour levels. The beam structure is
dominated by the imposed helical pattern with a characteristic wavelength of 
$\approx 3.0 R_b$ (to be compared with the value $\lambda = 3.5 R_b$ expected 
from the estimate of $\lambda$ in the previous paragraph) and an amplitude of 
$\approx 0.2 R_b$.

\subsection{Cocoon}
\label{ss:cocoon}

\begin{figure}
\plotone{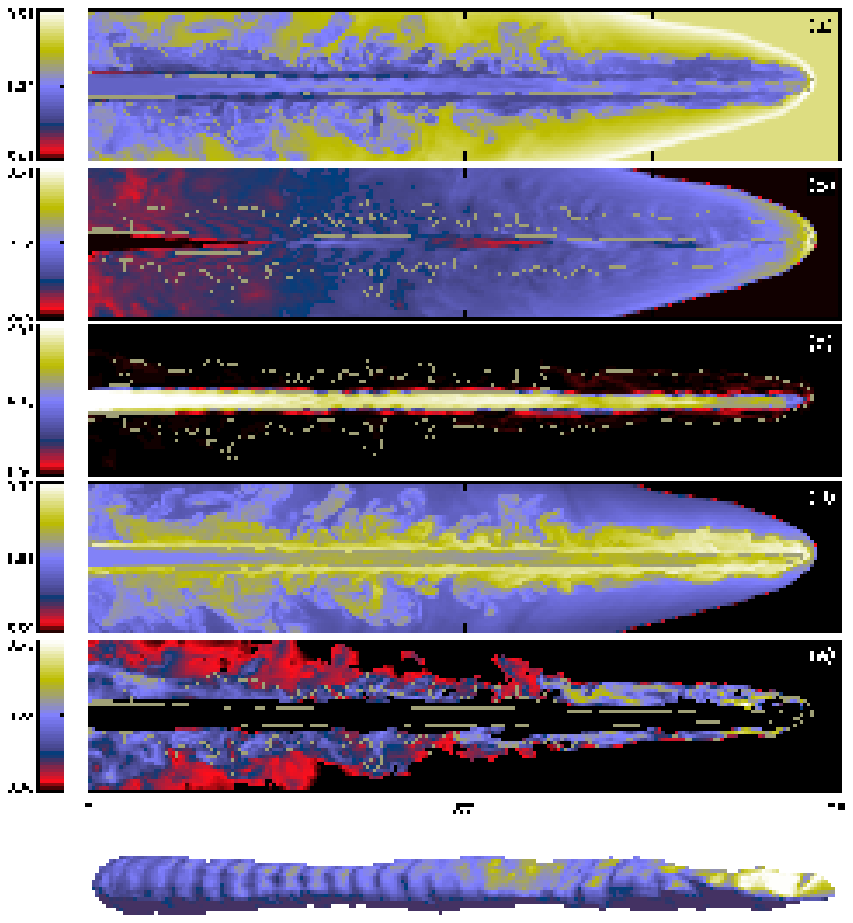}
\caption{Rest-mass density, pressure, flow Lorentz factor, specific
internal energy and backflow velocity distributions (from top to
bottom) of the model discussed in the text in the plane $y=0$ at the
end of the simulation.  White contour levels appearing in each frame
correspond to values of $f$ equal to 0.95 (inner contour; representative of 
the beam) and 0.05 (representative of the cocoon/shocked external medium
interface). The bottom panel displays the isosurface of $f=0.95$.
\label{f:n50p01a}}
\end{figure}

  The overall jet's morphology is characterizad by the presence of a highly
turbulent, subsonic, asymmetric cocoon. The pressure distribution outside the 
beam is nearly homogeneous giving rise to a symmetric bow shock 
(Fig.\,\ref{f:n50p01a}b). As in the classical case (Norman 1996), our 
relativistic 3D simulation shows less ordered structures in the cocoon. The 
cocoon remains quite thin ($\sim 2 R_b$) as long as the jet propagates 
efficiently.

  The flow field outside the beam shows that the high velocity backflow is
restricted to a small region in the vicinity of the hot spot
(Fig.\,\ref{f:n50p01a}e), the largest backflow velocities ($\sim 0.5c$) being 
significantly smaller than in 2D models. The flow with high Lorentz factor 
found in axisymmetric simulations (see flow patterns in Mart\'{\i} \etal 1996) 
appears here restricted to a thin layer around the beam and possesses
sub-relativistic speeds ($\sim 0.25c$). The magnitude of the backflow
velocities in the cocoon do not support relativistic beaming. 

\subsection{Beam and hot spot}
\label{ss:beam}

  Within the beam the perturbation pattern is superimposed to the
conical shocks at about 26 and 50\,$R_b$. The beam does not exhibit
the strong perturbations (deflection, twisting, flattening or even
filamentation) found by other authors (Norman (1996) for 3D classical
hydrodynamic jets; Hardee (1996) for 3D classical MHD jets). This can
be taken as a sign of stability, although it can be argued that our
simulation is not evolved far enough.  Obviously, the beam cross
section and the internal conical shock structure are correlated
(bottom panel in Figure\,\ref{f:n50p01a}). Before the first recollimation shock 
the beam cross section shrinks to an effective radius of $0.7 R_b$. After
this shock and in the rarefaction the beam reexpands and stretches due
to an elliptical surface mode (\eg Hardee 1996). Between $37R_b \la z
\la 50 R_b$ the beam flow is influenced by the second recollimation
shock, which causes a compression of the beam. A triangular mode seems
to grow in this region.

\begin{figure}
\plotone{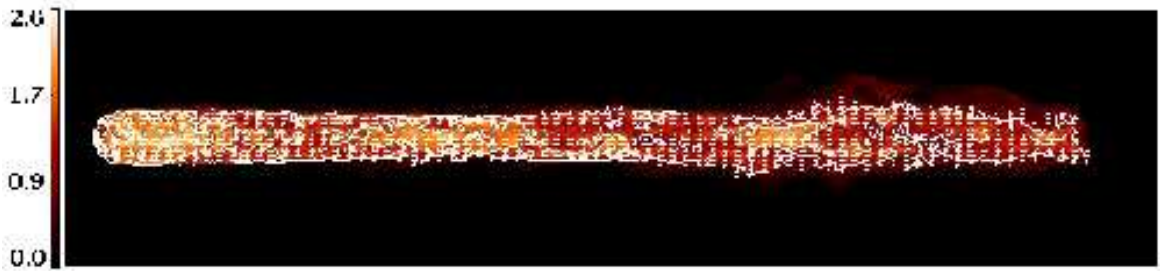}
\caption{Mean local apparent speed for the jet of Fig.\,1 observed at
an angle of 40 degrees. Arrows show the projected direction and
magnitude of the apparent motion the contours corresponding to values
of 1.0\,$c$, 1.6\,$c$, and 2.2\,$c$, respectively. Averages have been 
computed along the line of sight for each pixel in the image (computational 
cell) using the emission coefficient as a weight.
\label{f:apparent}}
\end{figure}

  The helical pattern propagates along the jet at nearly the beam
speed (see animation at http://scry.uv.es/aloy.html/JETS/videos/n50p01)
which could yield to superluminal components when viewed at
appropriate angles. Besides this superluminal pattern, the presence of
emitting fluid elements moving at different velocities and
orientations could lead to local variations of the apparent
superluminal motion within the jet.  This is shown in
Fig.\,\ref{f:apparent}, where we have computed the mean (along each
line of sight, and for a viewing angle of 40 degrees) local apparent
speed. The distribution of apparent motions is inhomogeneous and
resembles that of the observed individual features within knots in M87
(Biretta, Zhou, \& Owen 1995).

  The jet can be traced continuously up to the hot spot which
propagates as a strong shock through the ambient medium. Beam material
impinges on the hot spot at high Lorentz factors. We could not
identify a terminal Mach disk in the flow.  We find flow speeds near
(and in) the hot spot much larger than those inferred from the one
dimensional estimate. This fact was already noticed for 2D models by
Komissarov \& Falle (1996) and suggested by them as a plausible
explanation for an excess in hot spot beaming.

\subsection{Beam/cocoon shear layer}
\label{ss:shear}

  We find a layer of high specific internal energy
(Fig.\,\ref{f:n50p01a}d) surrounding the beam like in previous
axisymmetric models (Aloy \etal 1999).  A comparison with the
backflow velocities (Fig.\,\ref{f:n50p01a}e) shows that it is mainly
composed of forward moving beam material at a speed smaller than the
beam speed. The intermediate speed of the layer material is due to
shear in the beam/cocoon interface, which is also responsible for its
high specific internal energy.  The existence of such a boundary layer
has been invoked by several authors (Komissarov 1990, Laing 1996) to
interpret a number of observational trends in FRI radio sources. Swain, Bridle 
\& Baum (1998) have found evidence for these boundary layers in FRIIs (3C353).

\begin{figure}
\plotone{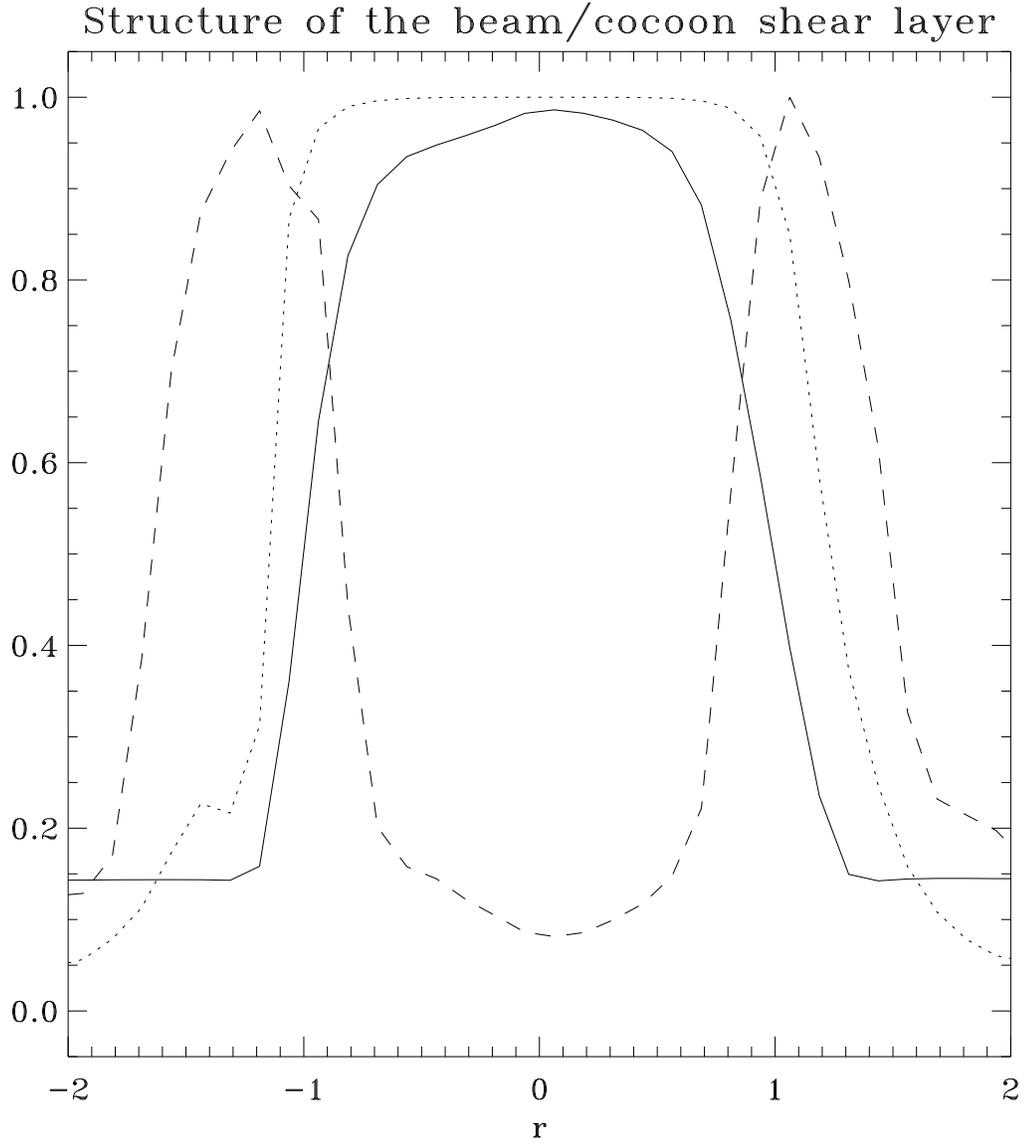}
\caption{Beam mass fraction (dotted line), flow Lorentz factor (filled line) and 
specific internal energy (dashed line), in arbitrary units, accross the beam 
($z=11.7 R_b$).
\label{f:shear_layer}}
\end{figure}

\begin{figure}
\plotone{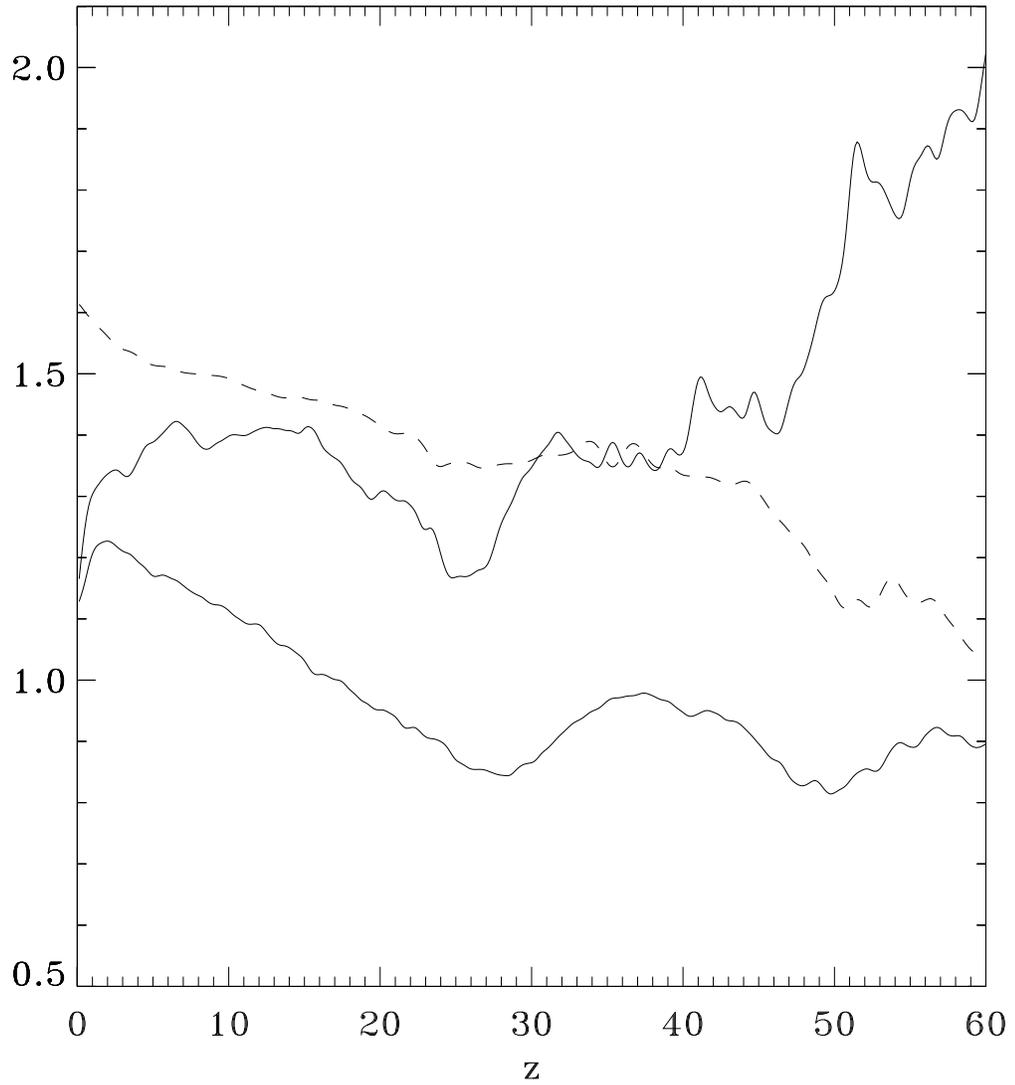}
\caption{Dashed line: Axial component of the momentum of the beam
particles (integrated accross the beam) along the jet axis at the end
of the simulation.  Solid lines: Mean beam radius along the axis for
$f \geq 0.2$ (top line) and $f \geq 0.8$ (bottom
line), respectively. Quantities are in code units.
\label{f:bound_cross_sect}}
\end{figure}

  The diffusion of vorticity caused by numerical viscosity is
responsible for the formation of the boundary layer. Although being
caused by numerical effects (and not by the physical mechanism of
turbulent shear) the properties of PPM--based difference schemes are
such that they can mimic turbulent flow to a certain degree (Porter \&
Woodward 1994).  Hence, our calculations represent a first approach to
study the development of shear layers in relativistic jets and their
observable consequences. The structure of both the shear layer and the beam 
core are sketched in Fig\,\ref{f:shear_layer}.  The specific internal energy of 
the gas in the shear layer (region with $0.2<f<0.8$) is typically more than
one order of magnitude larger than that of the gas in the beam
core. The shear layer broadens with distance from 0.2$R_b$ near the
nozzle to 1.1$R_b$ near the head of the jet
(Fig.\,\ref{f:bound_cross_sect}).

\subsection{Jet propagation efficiency and disruption}
\label{ss:propagation}
  
  From the head's position at the end of the simulation ($T=140.8$) a
mean jet advance speed of 0.47$c$ is obtained, but the jet's
propagation proceeds in two distinct phases: (i) for $t \la 100$ the
jet propagates roughly at the estimated 1D speed ($0.42c$); (ii) for
$t \ga 100$ the jet accelerates and propagates at a considerably
larger speed (0.55$c$). Comparing with the 3D simulation of Norman
(1996) we find a similar behaviour: after a short 1D phase and
before the deceleration, the jet transiently accelerates to a
propagation speed which is $\approx 20$\% larger than the
corresponding 1D estimate. This result contradicts the one obtained by
Nishikawa \etal (1997, 1998), who found a propagation speed of only
70\% of the corresponding 1D estimate in a shorter ($\approx$ 20 normalized
time units) simulation of a denser jet. Although the estimate do not take into
account the effect of an extra magnetic pressure in the external medium 
opposing to the jet propagation, in the case of Nishikawa \etal simulations
this magnetic pressure is negligible in comparison with the beam momentum 
density.
 
  Figure\,\ref{f:bound_cross_sect} shows the axial component of the
momentum of the beam particles (integrated across the beam) along the
axis, which decreases by 30\% within the first 60\,$R_b$.  Neglecting
pressure and viscous effects, and assuming stationarity the axial
momentum should be conserved, and hence the beam flow is
decelerating. The momentum loss goes along with the growth of the
boundary layer whose material is accelerated and heated by viscous
stresses.  Biconical shocks in the beam are responsible for the break
in the axial momentum profiles at $z=26 R_b$ and $z=50 R_b$, because
when the beam material passes a conical shock and enters into the
adjacent rarefaction fan, it is accelerated by local pressure
gradients.

  How can the jet accelerate while the beam material is decelerating?
Although the beam material decelerates, its terminal Lorentz factor is
still large enough to produce a fast jet propagation. On the other
hand, in 3D, the beam is prone to strong perturbations which can
affect the jet's head structure. In particular, a simple structure
like a terminal Mach shock will probably not survive when significant
3D effects develop. It will be substituted by more complex structures
in that case, \eg by a Mach shock which is no longer normal to the
beam flow and which wobbles around the instantaneous flow
direction. Another possibility is the generation of oblique shocks
near the jet head due to off--axis oscillations of the beam.  Although
difficult to check quantitatively (due to both the lack of an
operative definition for Mach disk identification and the present
resolution of our simulations) both possibilities will cause a less
efficient deceleration of the beam flow at least during some
epochs. At longer time scales the growth of 3D perturbations will
cause the beam to spread its momentum over a much larger area than
that it had initially, which will efficiently reduce the jet advance
speed.
  
\section{Conclusions}
\label{s:conc}

  We have presented a first attempt to analyze the morpho-dynamical
properties of 3D relativistic jets. From our simulations, we can
conclude that the coherent fast backflows found in axisymmetric models
are not present in 3D models. We have investigated the beam's response
to non-axisymmetric perturbations to check its stability.  During the
period of time studied by us ($t \la 150 R_b/c$), the beam does not
display the strong perturbations found by other authors in classical jets 
(Norman 1996,
Hardee 1996) and propagates according to the 1D estimate.  Small 3D
effects in the relativistic beam give rise to a lumpy distribution of
apparent speeds like that observed in M87 (Biretta, Zhou \& Owen
1995).  We have also analyzed the properties of the boundary layer
present in our model.

  Obviously, our study must be extended to a wider range of models and
perturbations.  In particular, stronger perturbations should be
considered to reach the nonlinear regime and to identify the acoustic
and mixing phases (Bodo 1998) leading to the jet disruption. Further
investigation also requires the dependence of the shear layer
properties on the perturbation parameters. Finally, appropriate
perturbations can be studied that mimic the wiggles observed in
specific sources both at pc (0836+710, Lobanov \etal 1998; 0735+178,
G\'omez \etal 1999) and kpc scales (M87; Biretta, Zhou \& Owen
1995). 
\bigskip 

\centerline{{ACKNOWLEDGEMENTS}}

This work has been supported in part by the Spanish DGES (grants
PB97-1164 and PB97-1432) and the CSIC-Max Planck Gesellschaft agreement. MAA 
expresses his gratitude to the
Conselleria d'Educaci\'o i Ci\`encia de la Generalitat Valenciana for
a fellowship. The calculations were carried out on two SGI Origin 2000
at the Centre Europeu de Paral.lelisme de Barcelona and at the Centre
de Inform\'{a}tica de la Universitat de Val\`encia.


\end{document}